\documentstyle[epsfig]{basi}
\begin{document}
\title[Spectrophotometry of comets]{Spectrophotometry of the comets C/2000 WM1 (LINEAR) and C/2002 C1 (Ikeya-Zhang)} 
\author[B. B. Sanwal et al.]%
       {B. B. Sanwal, Brijesh Kumar and Mahendra Singh \thanks{e-mail : sanwal, msing, brij@upso.ernet.in} \\ 
        State Observatory Manora Peak Nainital 263 129}
\maketitle
\label{firstpage}

\begin{abstract}
Spectrophotometric observations of the coma of the comets C/2000 WM1 (LINEAR) and C/2002 C1 (Ikeya-Zhang) were taken 
during Nov, Dec 2001 and Mar, Apr 2002 respectively  with 104-cm telescope of the State Observatory, Nainital. 
CN ($\lambda$ 3883 \AA) and C$_{2}$ swan bands ( $\lambda$ 4695, $\lambda$ 5165 and $\lambda$ 5538 \AA) have been 
identified in both the comets. Na I emission was detected in comet Ikeya-Zhang. An estimate of CN and C$_{2}$ abundances 
and their production rates have been derived. Dust production rates have also been determined.
\end{abstract}


\section{Introduction}
New comets come from the places in the solar system that are farthest and coldest and are believed to
contain matter that is unchanged since the formation of the solar system. Therefore the study of constituents of the 
comets is important for understanding the early stages of solar system formation.
At the center of the head of a comet is the nucleus. It is composed of chunks of matter. The most widely accepted theory of 
the comets, advanced in 1950 by Fred Whipple, is that the nucleus is "a dirty snowball". It constitutes ices of such 
molecules as water, carbon dioxide, ammonia, methane, with dust mixed in. This model explains many 
observed features of comets. The nucleus itself is too small to be observed directly from the earth. It 
is surrounded by coma which shines partly because of dust and gas reflecting sunlight.  
Gases liberated from the nucleus are excited by solar ultra-violet radiation and radiate it in 
the visible, adding to the visibility of comets. The spectrum of a comet shows sets of lines from simple molecules.
The comet is brightest and its tail is generally longest at the time of perihelion passage. However because of 
the orientation presented to the earth, it may not appear longest at this stage.
As the comet recedes from the sun, its tail fades.

\begin{table}[h]

\footnotesize
\caption{Basic data of the coma of the comets C/2000 WM1 (LINEAR) and C/2002 C1 (Ikeya-Zhang) at the time of observation }

\smallskip
\centering
\begin{tabular}{lcccccc} \hline
    Comet&        Date(UT)&  $\Delta$&    $r$&   $m_{1}$&               $\rho$&                      $D$\\
         &                &      (AU)&   (AU)&          &  ($\times 10^{4}$km)&                 (arcmin)\\ \hline  

         &                &          &       &          &                     &                         \\
C/2000 WM1& Dec 01.68, 2001&     0.317&  1.219&       5.9&               1.604&                     2.33\\
(LINEAR) & Dec 06.59, 2001&     0.329&  1.139&       5.7&               1.665&                     2.33\\
         &                &          &       &          &                     &                         \\
C/2002 C1& Mar 23.57, 2002&     0.733&  0.519&       4.0&                3.588&                     2.33\\
(Ikeya-Zhang)& Apr 15.92, 2002&     0.451&  0.815&       4.9&                2.208&                     2.33\\
         & Apr 16.92, 2002&     0.445&  0.830&       4.9&                2.188&                     2.33\\
         & Apr 18.92, 2002&     0.434&  0.864&       5.1&                2.124&                     2.33\\
         & Apr 29.90, 2002&     0.404&  1.048&       5.7&                1.977&                     2.33\\
         & Apr 30.90, 2002&     0.405&  1.066&       5.8&                1.982&                     2.33\\ \hline
\end{tabular} 

\leftskip 2.0cm
$\Delta$ = Geocentric distance; $r$ = Heliocentric distance\\
$m_{1}$ = Predicted integrated magntude; $\rho$ = Radius of circular region in the sky at $\Delta$\\
$D$ = Aperature diameter of the coma projected on the sky 
\end{table}

\begin{table}[h]
\caption{Observed fluxes in emission bands of the comets for the observed dates of Table 1.}

\smallskip
\centering
\footnotesize
\begin{tabular}{lccccc} \hline

      Comet&   $F$(C$_{2}$,$\Delta$v = 0)&   \multicolumn{4}{c}{$F$/$F$(C$_{2}$,$\Delta$v = 0)} \\ \cline{3-6}
           &$\times 10^{-10}ergs/cm^{2}/s$& CN($\Delta$v = 0)& C$_{2}$($\Delta$v = 1)& C$_{2}$($\Delta$v = 0)& CN($\Delta$v =$-1$)\\
           &          5165\AA&          3883\AA&             4695\AA&             5165\AA&     5538\AA \\ \hline

           &                 &                 &                    &                    &                     \\
  C/2000 WM1&            1.165&            0.892&               0.109&               1.000&                0.114\\
  (LINEAR) &            2.346&            0.803&               0.024&               1.000&                0.247\\
           &                 &                 &                    &                    &                     \\
C/2002 C1 &          139.306&            1.292&               0.302&               1.000&                0.369\\
(Ikeya-Zhang)&           62.486&            1.397&               0.008&               1.000&                0.379\\
           &           52.060&            1.265&               0.015&               1.000&                0.091\\
           &           27.598&            1.694&               0.146&               1.000&                0.615\\
           &           17.531&            1.756&               0.015&               1.000&                0.629\\
           &           00.548&            2.418&               0.058&               1.000&                1.000\\ \hline
\end{tabular}
\end{table}

\begin{table}[h]
\footnotesize
\caption{Column densities ($M$) and production rates ($Q$) of the comets for the observed dates of Table 1.}

\smallskip
\centering
\begin{tabular}{lccccccccc} \hline

      Comet&  \multicolumn{4}{c}{log($M$)}&&&  \multicolumn{3}{c}{log($Q$)} \\ \cline{2-5} \cline{7-10}
           &         CN&         C$_{2}$&        C$_{2}$&           C$_{2}$&  &  &           CN&        C$_{2}$& dust\\
  & ($\Delta$v = 0)&  ($\Delta$v = 1)& ($\Delta$v = 0)& $\Delta$v =$-1$)&  &  & ($\Delta$v = 0)& ($\Delta$v = 0)& 4850\AA\\ \hline

           &          &       &       &       & &   &       &        &             \\
  C/2000 WM1&    29.064& 28.347& 29.046& 28.429& &   & 24.735&  24.927&        10.209\\
   (LINEAR)&    29.294& 27.946& 29.322& 29.042& &   & 24.951&  25.192&        10.436\\
           &          &       &       &       & &   &       &        &             \\
 C/2002 C1&    31.165& 30.853& 31.108& 31.002& &   & 26.706&  26.555&        11.693\\
(Ikeya-Zhang)&    30.959& 28.889& 30.742& 30.636& &   & 26.523&  26.514&        11.625\\
           &    30.841& 29.117& 30.655& 29.941& &   & 26.404&  26.428&        11.265\\
           &    30.706& 29.825& 30.392& 30.509& &   & 26.281&  26.176&        10.784\\
           &    30.629& 28.756& 30.300& 30.427& &   & 26.202&  26.085&        10.848\\
           &    30.279& 28.841& 29.812& 30.139& &   & 25.848&  25.591&        10.636\\ \hline
\end{tabular}
\end{table}

The comet C/2000 WM1 (LINEAR) was detected by LINEAR team as apparently asteroid but was later found to be a comet 
by other observers. Later on Smithsonian Observatory astronomers (IAU Cir. No. 7546, 2001) successfully 
photographed the comet with 1.2-m reflector of Mount Hopkins, Arizona and at that time it's coma was 10 arcsec in size. 

On February 1, 2002 a comet C/2002 C1 was discovered visually by 
veteran comet hunter Kaoru Ikeya with a 25-cm aperature reflector 
and by Daqing Zhang with a home made 20-cm reflector. Ikeya has discovered four of the brightest comets of the past century, 
and this is his sixth comet. Zhang has been searching for comets for many years , but this is the first comet to carry 
his name. Marsden and Nakano (IAU Cir. No. 7843, 2002) have noted the similarity of the orbits of 
comets C/2002 C1 and C/1661 C1 and the
numerical integration of the 2002 orbits backwards yields a previous perihelion date within a couple of years of 1659, making 
the link rather likely. During March and April 2002, comet C/2002 C1 was observable to northern hemisphere observers low in 
the western sky. It moved northward in the sky till 25 April. It has been faintly visible to the naked eye 
from dark sky sites in March and remained a binocular object till late May or June. During March and April the comet 
showed an obvious tail which faded rapidly in April. Photographs taken on March 18, 2002 using CCD camera at the 
Cassegrain focus of 104-cm telescope of State Observatory Nainital showed well developed tail and emission of discrete jets.
The heliocentric distance on this date was 76 million km and the geocentric distance was 120 million km. The tail was extended 
upto about 0.8 million km and coma diameter was around 92 thousand km.

Extensive obervations of every comet are needed to determine production rates, column densities and surface brightnesses. 
Time resolved imaging of both the plasma tail and coma are needed to investigate mechanisms giving rise to the observed 
morphologies and associated changes. Therefore we took spectrophotometric observations of comets C/2000 WM1 (LINEAR) and 
C/2002 C2 (Ikeya-Zhang) during December 2001, and March, April 2002 respectively. 

\section{Observations and Data reductions}
The spectrophotometric observational system consists of a HR-320 spectrograph, a CCD detector, a detector 
interface and a computer. The spectrograph gives a dispersion of 2.4 \AA/pixel with a grating having 300 g/mm blazed 
at 5000 \AA. At the entrance slit of the spectrograph, a circular diaphram of  9 mm diameter corresponding to 2.33 arcmin 
as projected on the sky ($\rho$) and 
centered on the coma of the comet was used. The 1K $\times$ 1K CCD chip covers about 2500 \AA~ wide 
spectrum in a single exposure.

\begin{figure}[h]
\centerline{\psfig{file=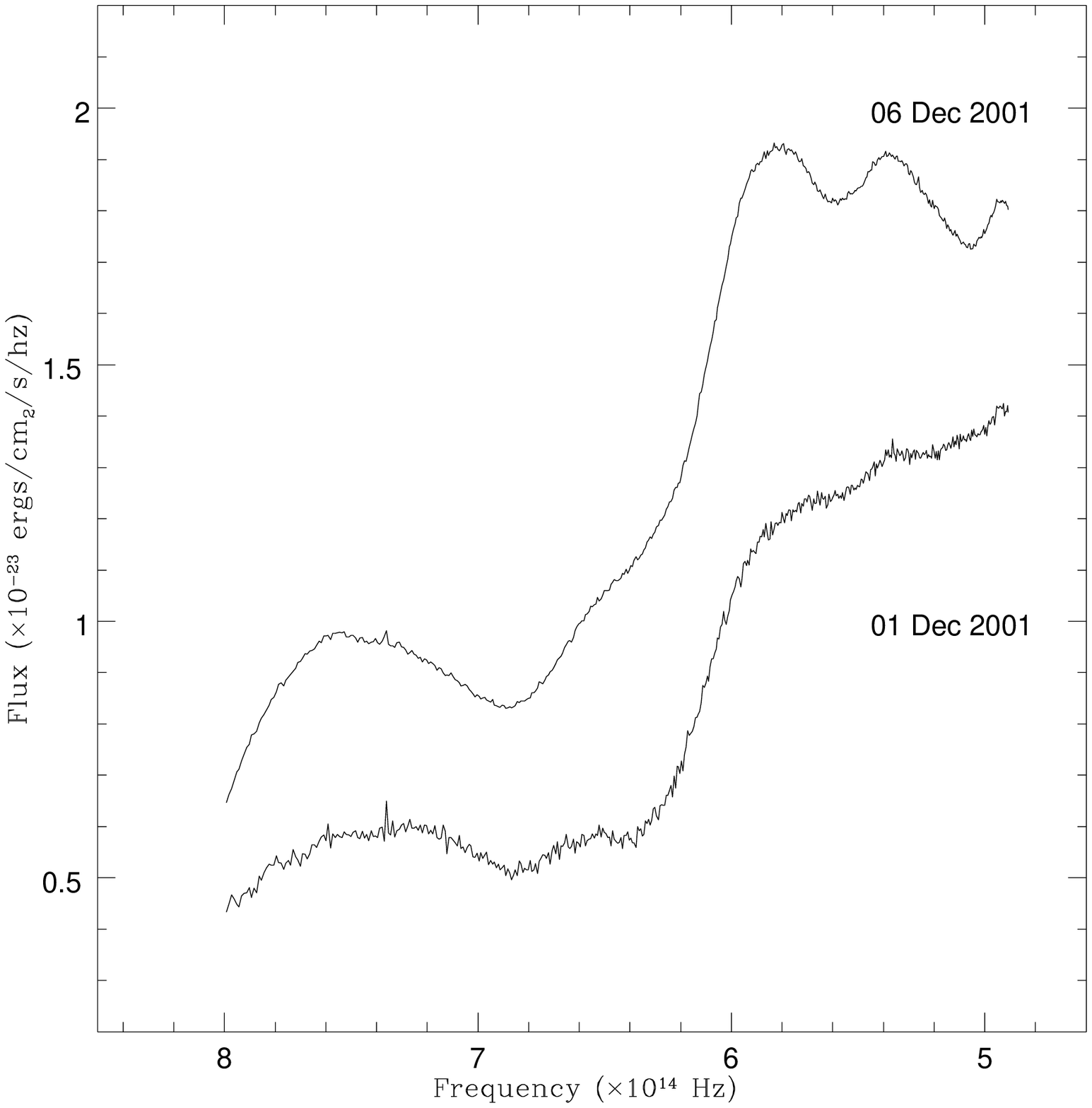,height=200pt,width=300pt}}
\caption{Absolute flux distribution of the head of the comet C/2000 WM1 (LINEAR)}
\end{figure}

At least three spectral frames of the each comet were obtained every night in the blue and red part of the spectrum. 
Along with the comet, standard stars were also observed to calibrate the flux of the comet spectra. Sufficient bias, 
twilight flats and sky frames were also taken. Data reduction is done using spectroscopic reduction software package of IRAF.
We have used bright A-type spectrophotomeric standards to calibrate wavelength of the spectra and hence our wavelength 
correction may have uncertainity of $\sim$ 10\AA. The flux and wavelength calibrated spectra (converted to frequency) are 
shown in Figures 1, 2 and 3. 
We have an usable range in spectrum from $\sim$ 3550 \AA~ to 6000 \AA~ for all dates. However on 15, 16, 18 
and 29 April 2002, we have observed in red region also upto 7000\AA~ and combined the spectra after wavelength and 
flux calibration. 
Comet C/2002 C1 is observed on six nights of Mar and Apr 2002 and comet C/2000 WM1 was observed on three nights of 
Nov and Dec 2001 (Nov 20, Dec 1 and 6). However, it was found to have no significant spectral features on Nov 20. 
Basic data at the time of observations for both the comets are given in Table 1. 

\begin{figure}[t]
\centerline{\psfig{file=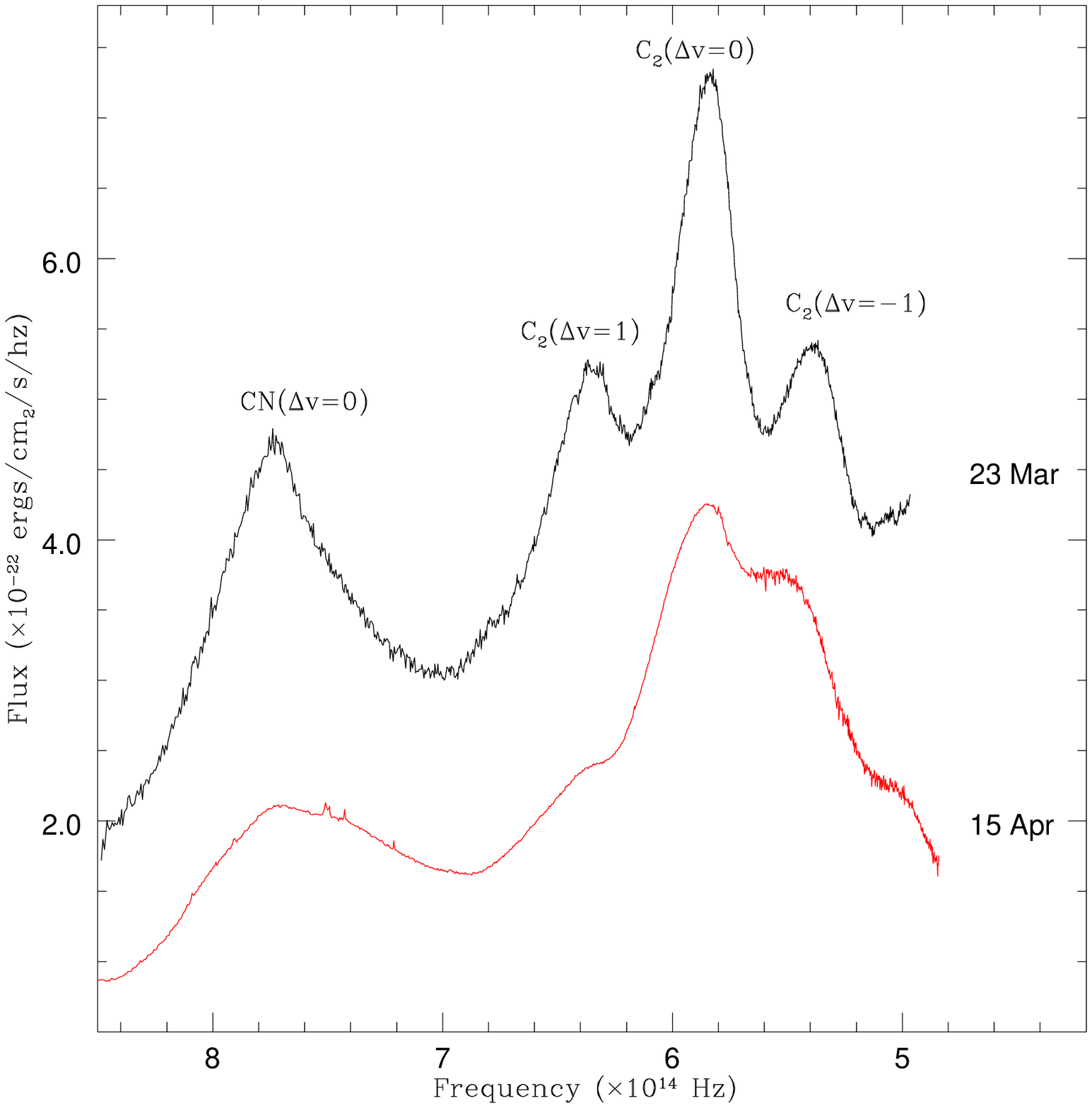,height=200pt,width=300pt}}
\caption{Absolute flux distribution of the head of the comet C/2002 C1 (Ikeya-Zhang)}
\end{figure}

\begin{figure}[h]
\centerline{\psfig{file=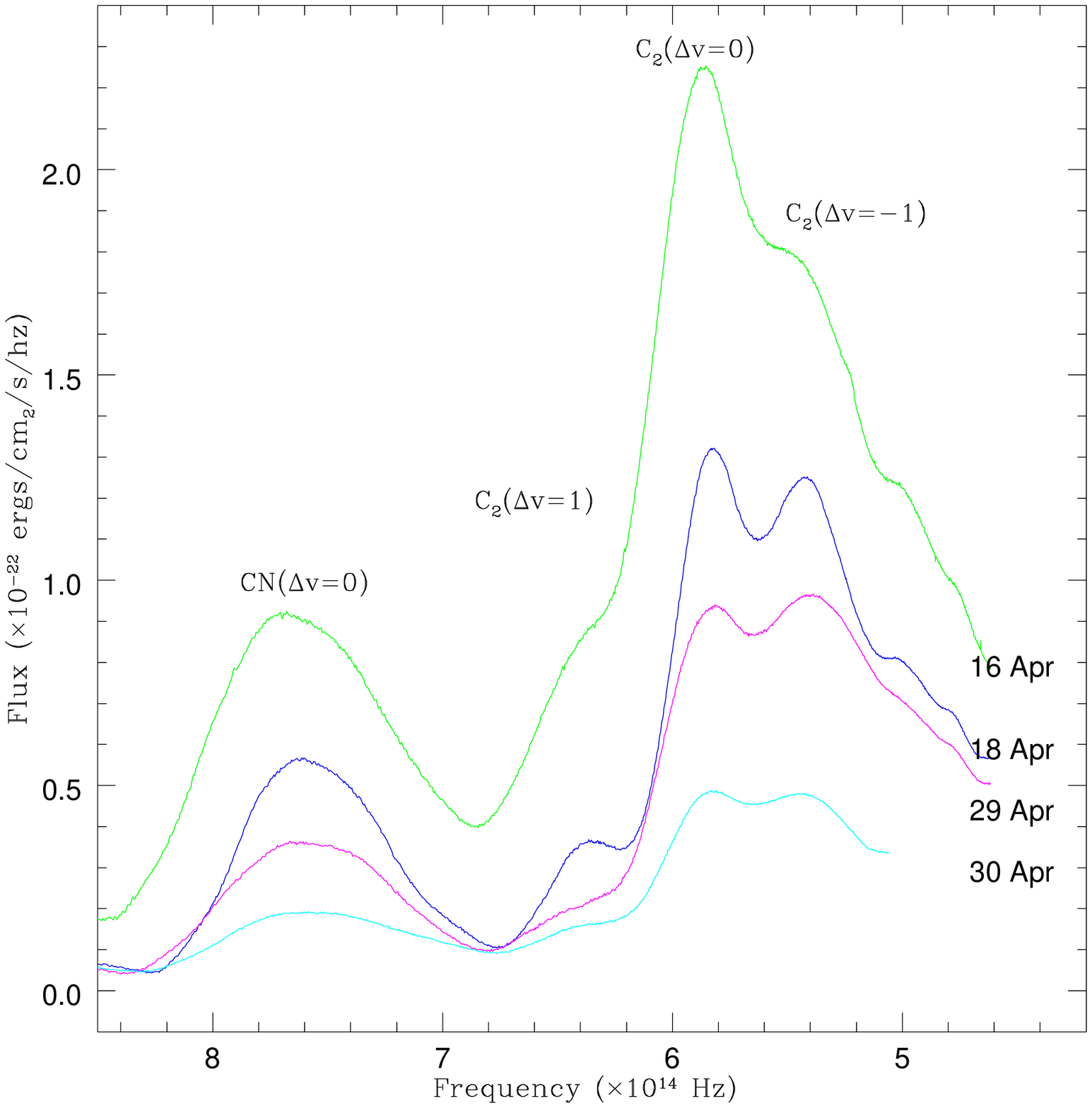,height=200pt,width=300pt}}
\caption{Absolute flux distribution of the head of the comet C/2002 C1 (Ikeya-Zhang)}
\end{figure}

\begin{figure}[h]
\centerline{\psfig{file=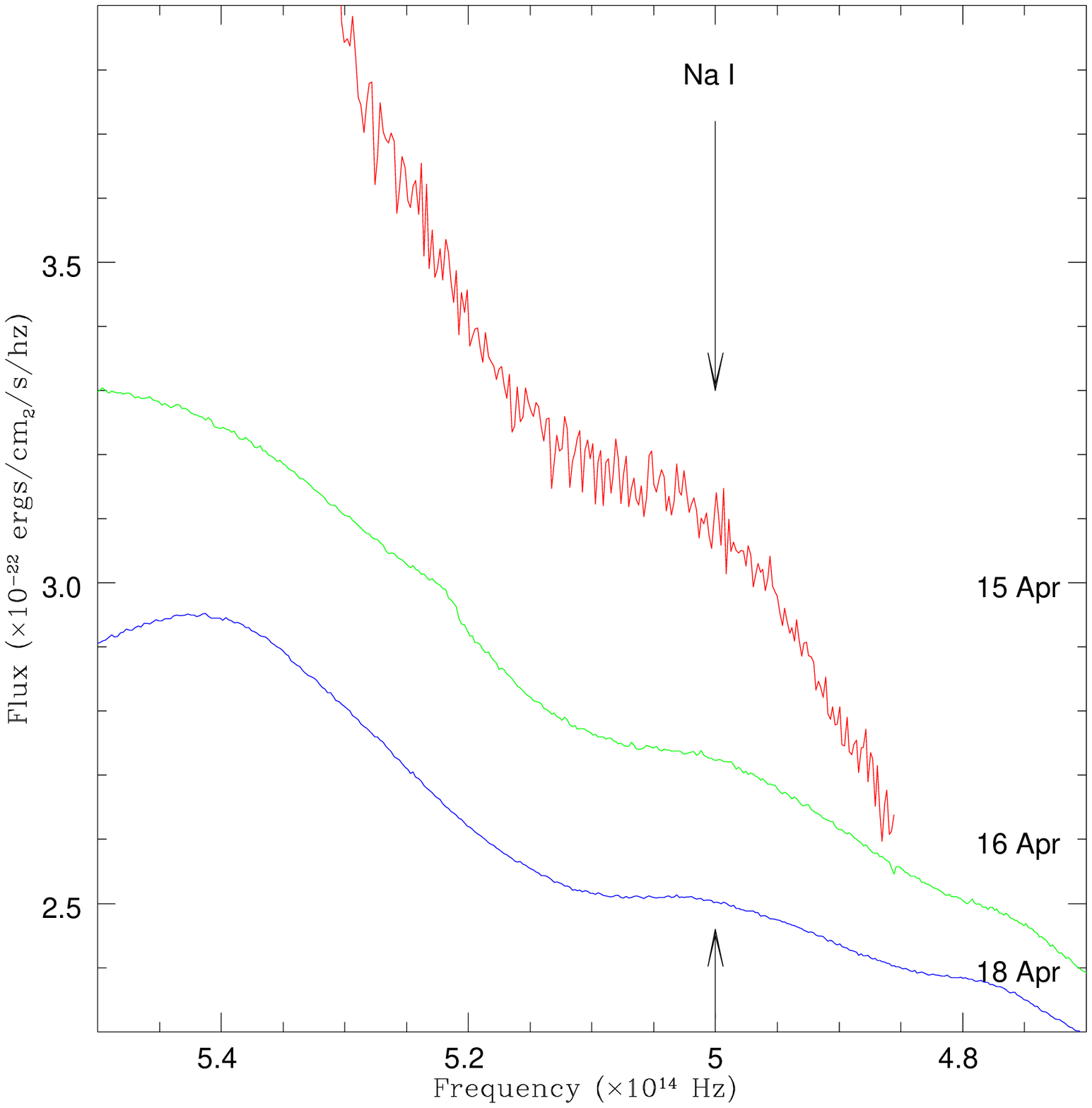,height=200pt,width=300pt}}
\caption{The extended portion in the Na I emission region for the comet C/2002 C1. } 
\end{figure}

\section{Emission bands, column densities and production rates}
The prominent features, as can be seen in Figures 1, 2 and 3 
are CN($\Delta$v = 0) at 3883 \AA, C$_{2}$($\Delta$v $= +1, 0, -1$) 
at 4695, 5165 and 5538 \AA~ respectively. These features are weaker for C/2000 WM1. 
In order to measure fluxes in these emission bands, the continuum in spectrum
was located by selecting wavelengths 3600 \AA, 4350 \AA, 4850\AA, 5400\AA~ and 5800\AA~ which are free from emission bands. 
The area of strong emission bands were measured and converted
into the total flux. Emission band fluxes relative to C$_{2}$ (5165 \AA) is given in Table 2.

The number of molecules of each species, 
contained in a cylinder of radius defined by the diaphram used, and extending entirely 
through the coma was evaluated using the expression (Millis et. al. 1982) which is given as, 

\begin{equation}
 \log M(\rho) = \log F(\rho) + 27.449 + 2\log(\Delta r) - \log g
\end{equation}

\noindent where $F$ is the observed flux in cgs units, $\rho$ is the projected radius of circular region in the sky at $\Delta$, 
$r$ and $\Delta$ are the heliocentric and geocentric distances of the comet
respectively in AU and $g$ the fluorescence efficiency (in cgs units) per molecule at 1 AU. We used the values of fluorescence
efficiency for C$_{2}$ and C$_{3}$  from Sivaraman et al (1987). Because of the swing effect $g$ varies significantly for
CN with the comets heliocentric radial velocity. In order to determine radial velocity, the orbital elements for C/2000 WM1 
and C/2002 C1 were taken from British Astron. Assoc. Cir. No.780, 2001 and British Astron. Assoc. Cir. No. 782, 2001 
respectively. During our observations the radial velocity 
varied from 13.5 km s$^{-1}$ to 30.0 km s$^{-1}$ for C/2002 C1 and was $\sim$ 28 km s$^{-1}$ for the C/2000 WM1. The 
value of $g$ was obtained for all the values of radial velocities from the Figure of Tatum and Gillespie (1977). 
Therefore the variation of $g$ has been incorporated in the calculation of column densities of the molecule CN. The column 
densities thus obtained are listed in Table 3. 

The column densities thus calculated were converted into production rates ($Q$), assuming a Haser model, through the 
relationship given by A'Hearn and Cowan (1975),

\begin{equation}
 M(\rho) = Q V^{-1} \rho \left[ \int_{x}^{\mu x} K_{0}(y) dy + (1/x) (1 - 1/\mu) + K_{1}(\mu x) - K_{1}(x) \right]
\end{equation}

\noindent where $V$ is the velocity of released species; $x$ is the ratio between $\rho$ and daughter molecule scale lengths; 
$\mu$ is the ratio between daughter and parent molecules scale 
lengths; $K_{0}$ and $K_{1}$ are modified Bessel functions of the second kind of order 0 and 1. Following Delsemme (1982) we 
assumed $V = 0.58/\sqrt r$.  The parent and daughter molecular scale lengths are taken from Cocharan (1985). The 
resulting production rates are given in Table 3. 

Production rates for dust have been estimated on the simplest possible model, spherically 
symmetric, uniform outflow ignoring variations in particle sizes, scattering angle, etc. The relationship derived 
by A'Hearn and Cowan (1975) is used to evaluate the production rate, $Q$, of dust in arbitrary units.

\begin{equation}
Q = L r^{2}/\rho
\end{equation}

\noindent where $L$ is the luminosity of the comet at $\lambda = 4850$ \AA. The dust production rates evaluated are listed 
in Table 3.

\begin{figure}[t]
\centerline{\psfig{file=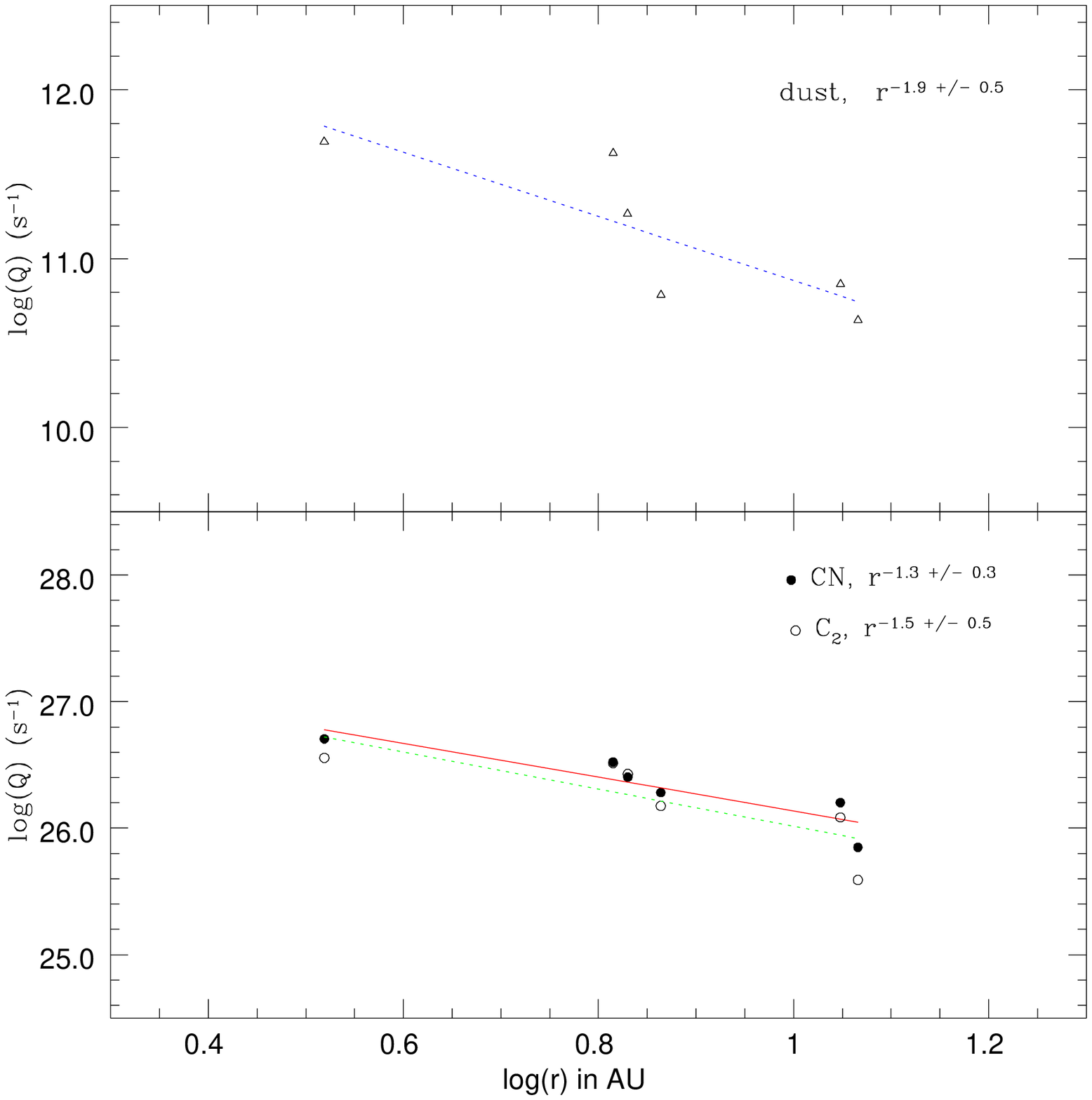,height=200pt,width=300pt}}
\caption{The production rates of molecular species CN, C$_{2}$ and dust as a function of heliocentric distance for the comet 
C/2002 C1. In the lower panel solid line represent CN and the dotted line represent C$_{2}$}
\end{figure}

\section{Discussions}
The prominent emission features as seen in Figure 1, 2 and 3 
are CN($\Delta v=0$) at 3888 \AA~ and C$_{2}$ ($\Delta v = +1,0,-1$) at 4690, 5160 and 5530 \AA. The strongest feature in the 
whole spectrum is due to C$_{2}$ ($\Delta v = 0$) at 5167 \AA. Figure 4 shows the extended region of the C/2002 C1 spectra 
around Na I emission at 5893 \AA, which has been detected also in the earlier observation (IAU Cir. No. 7914, 2002). However, 
Na I emission line was not taken into account for the purpose of calculating the parameters like production rates and 
column densities, as there was not sufficient signal at that wavelength.
C/2002 C1 was nearest to the sun on March 23. Therefore the activity was maximum on that date. This 
is reflected in the continuum and in the production rates of molecules and dust which is maximum on March 23. 
As the heliocentric distance increased the activity was decreased.

Figure 5 shows the production rates as a function of 
heliocentric distance for the comet C/2002 C1 and it varies approximately as $r^{-2}$ for C$_{2}$, CN and dust 
which is in agreements with the classical model of equilibrium vaporization of the nucleus. 
In the case of C/2000 WM1 (LINEAR), the emission bands are weak although its total integrated magnitude is brighter.
This may be due to its large heliocentric distance (at the time of observations) and the contribution
of dust production is higher in comparison to others.
The weak emission bands observed in C/2000 WM1 inspite of its bright integrated magnitude is probably 
because of the comets large heliocentric distance and because of much more dust content in comparison to other comets.

\section*{Acknowledgements}
The authors are thankful to the anonymous referee for his helpful comments.


\label{lastpage}
\end{document}